\documentclass[11pt]{article}
\usepackage[russian]{babel}
\usepackage{amssymb,amsmath,amsthm}
\newcommand{\bc}{\begin{center}}
\newcommand{\ec}{\end{center}}
\newcommand{\be}{\begin{equation}}
\newcommand{\ee}{\end{equation}}
\newcommand{\ba}{\begin{array}}
\newcommand{\ea}{\end{array}}
\newcommand{\edc}{\end{document}}

\begin{document}
УДК 517.98
\begin{center}
\subsection*{Трансляционно-инвариантность периодических мер Гиббса для модели Поттса на дереве Кэли}
\end{center}

\begin{center}
Р.М.Хакимов\footnote{Наманганский государственный университет, Наманган, Узбекистан.\\
E-mail: rustam-7102@rambler.ru},
М.Т.Махаммадалиев\footnote{Наманганский государственный университет, Наманган, Узбекистан.\\
E-mail: mmtmuxtor93@mail.ru}
\end{center}\

Работа является продолжением работы \cite{RK}. Изучается модель Поттса с нулевым внешним полем на дереве Кэли.
Для антиферромагнитной модели Поттса с $q$-состояниями на дереве
Кэли порядка два и для ферромагнитной модели Поттса с $q$-состояниями на дереве
Кэли порядка $k$ показано, что при любых значениях параметра все периодические
меры Гиббса являются трансляционно-инвариантными.\\

\textbf{Ключевые слова}: дерево Кэли, конфигурация, модель Поттса,
мера Гиббса, периодические меры, трансляционно-инвариантные меры.

\section{Введение}\

Решения проблем, возникающих в результате исследований при изучении
термодинамических свойств физических и биологических систем, в основном
приводятся к задачам теории мер Гиббса. Известно, что каждой предельной
мере Гиббса сопоставляется одна фаза физической системы.
Модель Поттса является обобщением модели Изинга, которая хорошо
изучена на решетке $Z^d$ и на дереве Кэли. Понятие меры Гиббса
для модели Поттса на дереве Кэли вводится обычным образом (см. \cite{6}-
\cite{R}).

В работе \cite{Ga8} изучена ферромагнитная модель Поттса с тремя
состояниями на дереве Кэли второго порядка и показано
существование критической температуры $T_c$ такой, что при $T<T_c$
существуют три трансляционно-инвариантных и несчетное число не
трансляционно-инвариантных мер Гиббса. В работе \cite {GN}
обобщены результаты работы \cite{Ga8} для модели Поттса с конечным
числом состояний на дереве Кэли произвольного (конечного) порядка.

Показано, что на дереве Кэли произвольного порядка
трансляционно-инвариантная мера Гиббса антиферромагнитной модели
Поттса с $q$-состояниями и с внешним полем единственна (см.\cite{R}, стр.109, Теорема 5.2).
Работа \cite{Ga13} посвящена модели Поттса со счетным числом состояний и
c ненулевым внешним полем и доказано, что эта модель имеет
единственную трансляционно-инвариантную меру Гиббса.

В работе \cite{RK} изучены периодические меры Гиббса и при некоторых условиях доказано, что все
периодические меры Гиббса являются трансляционно-инвариантными. В частности, для ферромагнитной
модели Поттса с \emph{тремя} состояниями на дереве Кэли прозвольного порядка и для антиферромагнитной модели Поттса
с \emph{тремя} состояниями на дереве Кэли второго порядка \emph{при некоторых условиях}, показано, что все периодические меры Гиббса являются трансляционно-инвариантными. Кроме того, найдены условия, при которых модель Поттса с ненулевым внешним полем имеет периодические меры Гиббса. Работа \cite{KhR1} является продолжением работы \cite{RK}. Доказана существование не менее трех периодических мер Гиббса с периодом два на дереве Кэли порядка
три и четыре для модели Поттса с тремя состояниями и с нулевым внешним полем. А в работе \cite{KhR} изучена модель Поттса с $q$-состояниями на дереве Кэли порядка $k\geq3$ и на некоторых инвариантах показано существование периодических (не трансляционно-инвариантных) мер Гиббса при некоторых условиях на параметры этой модели. Кроме того, указана нижняя граница количества существующих периодических мер Гиббса. В работе \cite{KRK} дано полное описание трансляционно-инвариантных мер Гиббса для ферромагнитной модели Поттса с $q$-состояниями и показано, что
их количество равно $2^q-1$, а в работе \cite{KR} изучена задача крайности этих мер. В работе \cite{KhH} улучшены результаты из \cite{KhR} и даны явные формулы для трансляционно-инвариантных мер Гиббса для модели Поттса с тремя состояниями на дереве Кэли порядка $k=3$.

В данной работе обобщены некоторые результаты из \cite{RK}: Для антиферромагнитной
модели Поттса с $q$-состояниями на дереве Кэли порядка два показано, что при любых значениях параметра все периодические меры Гиббса являются трансляционно-инвариантными. А также для ферромагнитной модели Поттса с $q$-состояниями на дереве Кэли порядка $k$ показано, что все периодические
меры Гиббса являются трансляционно-инвариантными.

\section{Определения и известные факты.}\

Дерево Кэли $\Im^k$ порядка $ k\geq 1 $ - бесконечное дерево, т.е.
граф без циклов, из каждой вершины которого выходит ровно $k+1$
ребер. Пусть $\Im^k=(V,L,i)$, где $V$ --- есть множество вершин
$\Im^k$, $L$ --- его множество ребер, и $i$ --- функция
инцидентности, сопоставляющая каждому ребру $l\in L$ его концевые
точки $x, y \in V$. Если $i (l) = \{ x, y \} $, то $x$ и $y$
называются  {\it ближайшими соседями вершины} и обозначается $l =
\langle x,y\rangle $. Расстояние $d(x,y), x, y \in V$ на дереве Кэли
определяется формулой
$$
d (x, y) = \min \ \{d | \exists x=x_0, x_1,\dots, x _ {d-1},
x_d=y\in V \ \ \mbox {такой, что} \ \ \langle x_0, x_1\rangle,\dots, \langle x _
{d-1}, x_d\rangle\} .$$

Для фиксированого $x^0\in V$ обозначим $ W_n = \ \{x\in V\ \ | \ \
d (x, x^0) =n \}, $
$$ V_n = \ \{x\in V\ \ | \ \ d (x, x^0) \leq n \},\ \
L_n = \ \{l = \langle x,y\rangle \in L \ \ | \ \ x, y \in V_n \}. $$

Мы рассмотрим модель, где спиновые переменные принимают значения
из множества $\Phi = \ \{1, 2,\dots, q \},$ $ q\geq 2 $ и
расположены на вершинах дерева. Тогда \emph{ конфигурация} $\sigma$ на
$V$ определяется как функция $x\in V\to\sigma (x) \in\Phi$; множество
всех конфигураций совпадает с $\Omega =\Phi ^ {V} $.

Гамильтониан модели Поттса определяется как
$$H(\sigma)=-J\sum_{\langle x,y\rangle\in L}
\delta_{\sigma(x)\sigma(y)}-\alpha\sum_{x\in V}
\delta_{1\sigma(x)},\eqno(1)$$ где $J\in R$, $\alpha\in R-$
внешнее поле, $\langle x,y\rangle-$ ближайшие соседи и
$\delta_{ij}-$ символ Кронекера:
$$\delta_{ij}=\left\{\begin{array}{ll}
0, \ \ \mbox{если} \ \ i\ne j\\[2mm]
1, \ \ \mbox{если} \ \ i= j.
\end{array}\right.
$$
Определим конечномерное распределение вероятностной меры $\mu_n$ в
обьеме $V_n$ как $$\mu_n(\sigma_n)=Z_n^{-1}\exp\left\{-\beta
H_n(\sigma_n)+\sum_{x\in W_n}\tilde h_{\sigma(x),x}\right\},\eqno(2)$$
где $\beta=1/T$, $T>0$--температура,  $Z_n^{-1}$ нормирующий
множитель и $\{\tilde h_x=(\tilde h_{1,x},\dots, \tilde h_{q,x})\in R^q, x\in V\}$
совокупность векторов и
$$H_n(\sigma_n)=-J\sum_{\langle x,y\rangle\in L_n}
\delta_{\sigma(x)\sigma(y)}-\alpha\sum_{x\in V_n}
\delta_{1\sigma(x)}.$$

Говорят, что вероятностное распределение (2) согласованное, если
для всех $n\geq 1$ и $\sigma_{n-1}\in \Phi^{V_{n-1}}$:
$$\sum_{\omega_n\in \Phi^{W_n}}\mu_n(\sigma_{n-1}\vee
\omega_n)=\mu_{n-1}(\sigma_{n-1}).$$

Здесь $\sigma_{n-1}\vee \omega_n$  есть объединение конфигураций.
В этом случае, существует единственная мера $\mu$ на $\Phi^V$
такая, что для всех $n$ и $\sigma_n\in \Phi^{V_n}$
$$\mu(\{\sigma|_{V_n}=\sigma_n\})=\mu_n(\sigma_n).$$
Такая мера называется расщепленной гиббсовской мерой,
соответствующей гамильтониану (1) и векторнозначной функции $\tilde h_x,
x\in V$. Следующее утверждение описывает условие на $\tilde h_x$, обеспечивающее
согласованность $\mu_n(\sigma_n)$.

\textbf{Теорема 1}.\cite{1} \textit{Вероятностное распределение
$\mu_n(\sigma_n)$, $n=1,2,\ldots$ в (2) является согласованной
тогда и только тогда}, \textit{когда для любого} $x\in V$
\textit{имеет место следующее  $$h_x=\sum_{y\in
S(x)}F(h_y,\theta,\alpha),\eqno(3)$$ где $F: h=(h_1,
\dots,h_{q-1})\in R^{q-1}\to
F(h,\theta,\alpha)=(F_1,\dots,F_{q-1})\in R^{q-1}$ определяется
как:
$$F_i=\alpha\beta\delta_{1i}+\ln\left({(\theta-1)e^{h_i}+\sum_{j=1}^{q-1}e^{h_j}+1\over
\theta+ \sum_{j=1}^{q-1}e^{h_j}}\right),$$
$\theta=\exp(J\beta)$, $S(x)-$ множество прямых потомков точки
$x$ и
$h_x=\left(h_{1,x},\dots,h_{q-1,x}\right)$ с условием
$$
h_{i,x}={\tilde h}_{i,x}-{\tilde h}_{q,x}, \ \ i=1,\dots,q-1.$$}\

Известно, что существует взаимнооднозначное соответствие между
множеством $V$ вершин дерева Кэли порядка $k\geq 1 $ и группой $G
_{k},$ являющейся свободным произведением $k+1$ циклических групп
второго порядка с образующими $a_1, a_2,\dots, a_{k+1} $,
соответственно.\

Пусть $\widehat{G}_k-$ нормальный делитель конечного индекса группы $G_k$.

\textbf{Определение 1}. Совокупность векторов $h=\{h_x,\, x\in G_k\}$
называется $ \widehat{G}_k$-периодической, если  $h_{yx}=h_x$ для
$\forall x\in G_k, y\in\widehat{G}_k.$

$G_k-$ периодические совокупности называются трансляционно-инвариантными.

\textbf{Определение 2}. Мера $\mu$ называется
$\widehat{G}_k$-периодической, если она соответствует
$\widehat{G}_k$-периодической совокупности векторов $h$.

В \cite{RK} доказана следующая теорема.

\textbf{Теорема 2.} \cite{RK} \textit{Пусть $H-$ нормальный делитель
конечного индекса в $G_k.$ Тогда для модели Поттса все $H-$
периодические меры Гиббса являются либо $G_k^{(2)}-$
периодическими, либо трансляционно-инвариантными, где $G_k^{(2)}$
есть подгруппа, состоящая из слов четной длины.}

\section{Антиферромагнитный случай.}\

Рассмотрим случай $q\geq 3, \ \alpha=0$, т.е. $\sigma:V\rightarrow\Phi=
\{1,2,3,...,q\}$. В силу Теоремы 2 имеются только
$G^{(2)}_k$-периодические меры Гиббса, которые соответствуют
совокупности векторов $h=\{h_x\in R^{q-1}: \, x\in G_k\}$ вида
$$h_x=\left\{%
\begin{array}{ll}
    h, \ \ \ $ если $ |x|-\mbox{четно} $,$ \\
    l, \ \ \ $ если $ |x|-\mbox{нечетно} $.$ \\
\end{array}%
\right. $$
 Здесь $h=(h_1,h_2,...,h_{q-1}),$ $l=(l_1,l_2,...,l_{q-1}).$
Тогда в силу (3) имеем:
$$
\left\{%
\begin{array}{ll}
    h_{i}=k\ln{(\theta-1)\exp(l_i) + \sum_{j=1}^{q-1}exp({l_j})+1\over \sum_{j=1}^{q-1}exp({l_j})+\theta},\\[3 mm]
    l_{i}=k\ln{(\theta-1)\exp(h_i) + \sum_{j=1}^{q-1}exp({h_j})+1\over \sum_{j=1}^{q-1}exp({h_j})+\theta},  \\
\end{array}%
i=\overline{1,q-1}. \right.$$\

Введем следующие обозначения: $\exp(h_i)=x_i,\ \exp(l_i)=y_i.$
Тогда последнюю систему уравнений при $i=\overline{1,q-1}$ можно
переписать:
$$
\left\{%
\begin{array}{ll}
    x_{i}=\left({(\theta-1)y_i + \sum_{j=1}^{q-1}y_j+1\over \sum_{j=1}^{q-1}y_j+\theta}\right)^k, \\[3 mm]
    y_{i}=\left({(\theta-1)x_i + \sum_{j=1}^{q-1}x_j+1\over \sum_{j=1}^{q-1}x_j+\theta}\right)^k.\\
    \end{array}%
\right.\eqno(4)$$\

Рассмотрим отображение $W:R^{q-1}\times R^{q-1} \rightarrow
R^{q-1}\times R^{q-1},$ определенное
следующим образом:
$$
\left\{%
\begin{array}{ll}
    x_{i}^{'}=\left({(\theta-1)y_i + \sum_{j=1}^{q-1}y_j+1\over \sum_{j=1}^{q-1}y_j+\theta}\right)^k, \\[3 mm]
    y_{i}^{'}=\left({(\theta-1)x_i + \sum_{j=1}^{q-1}x_j+1\over \sum_{j=1}^{q-1}x_j+\theta}\right)^k.\\
    \end{array}%
\right.\eqno(5)$$\

Заметим, что (4) есть уравнение $z=W(z).$ Чтобы
решить систему уравнений (4), надо найти неподвижные точки
отображения (5): $z^{'}=W(z)$, где $z=(x_1,...,x_{q-1},y_1,...,y_{q-1})$.\

\textbf{Лемма 1.}\textit{Следующие множества являются инвариантными относительно отображения $W$:}
$$I_1=\{z\in R^{2q-2}: x_1=x_2=\ldots =x_{q-1}=y_1=y_2=\ldots =y_{q-1}\},$$
$$I_2=\{z\in R^{2q-2}: x_1=x_2=\ldots =x_{q-1}, \ y_1=y_2=\ldots =y_{q-1}\},$$
$$I_3=\{z\in R^{2q-2}: x_i=y_i, \ i=1,2,\ldots, q-1\},$$
$$I_4=\{z\in R^{2q-2}: x_i=y_{q-i}, \ i=1,2,\ldots, q-1\}\},$$
$$I_5=\{z\in R^{2q-2}: x_1=y_1=1\}, \ \ I_6=\{z\in R^{2q-2}: x_{q-1}=y_{q-1}=1\}.$$

Доказывается аналогично Лемме 2 из \cite{RK}.

\textbf{Замечание 1.} Заметим, что отображение $W$ может иметь инвариантные множества, отличные от $I_1-I_6$, т.е. множества $I_1-I_6$ не полностью описывают все инвариантные множества отображения $W$.

\textbf{Лемма 2.} \textit{Меры Гиббса для модели Поттса на инвариантных множествах $I_1$  и $I_3$  являются трансляционно-инвариантными.}

Доказательство очевидно, так как на инвариантных множествах $I_1$  и $I_3$ имеем $h_x=const$.

\textbf{Замечание 2.} 1) При $q=2$ модель Поттса совпадает с
моделью Изинга, которая была изучена в работе \cite{1}.

2) В случае $k=2, \ q=3$, $J<0$ и $\alpha=0$ было доказано, что на
инвариантных множествах $I_1-I_6$ все $G_k^{(2)}$-периодические меры Гиббса являются
трансляционно-инвариантными (см. \cite{RK}).

Справедлива следующая теорема.

\textbf{Теорема 3}. Пусть $k=2, \ q\geq2, \ J<0, \ \alpha=0$. Тогда для модели Поттса
$G_k^{(2)}$-периодическая мера Гиббса единственна. Более того, эта мера совпадает с единственной трансляционно-инвариантной мерой Гиббса.

\textbf{Доказательство.} Заметим, что $x_i=y_i=1, \ i=1,2,\ldots
,q-1$ является решением системы уравнений (4), которая состоит из
$(2q-2)$ уравнений. Покажем, что (4) не имеет других решений. Для
этого в (4) выражения для $y_i$ подставим в правые части первых
$q-1$ уравнений. Тогда получим
$$
\left\{%
\begin{array}{ll}
    \sqrt{x_1}=\frac{\theta \tilde x_1+\tilde x_2+ \tilde x_3 +\ldots+\tilde x_{q-1}+\gamma}{\tilde x_1+ \tilde x_2 +\tilde x_3 +\ldots+\tilde x_{q-1}+\theta\gamma}, \\[3 mm]
    \sqrt{x_2}=\frac{\theta \tilde x_2 + \tilde x_1+\tilde x_3+\ldots+\tilde x_{q-1}+\gamma}{\tilde x_1+\tilde x_2+\tilde x_3 +\ldots+\tilde x_{q-1}+\theta\gamma},\\[3 mm]
    \sqrt{x_3}=\frac{\theta \tilde x_3+\tilde x_1+\tilde x_2 + \ldots+\tilde x_{q-1} +\gamma}{\tilde x_1+\tilde x_2+\tilde x_3+\ldots+\tilde x_{q-1}+\theta\gamma},\\[3 mm]
    \ldots \ldots \ldots \ldots \ldots \ldots \ldots \ldots \ldots \\[3 mm]
    \sqrt{x_{q-1}}=\frac{\theta\tilde  x_{q-1}+\tilde x_1+\tilde x_2+\ldots+\tilde x_{q-2}+\gamma}{\tilde x_1+\tilde x_2+\tilde x_3+\ldots+\tilde x_{q-1}+\theta\gamma},\\
    \end{array}%
\right.\eqno(6)$$\
где $\tilde x_1=(\theta x_1+x_2+x_3+\ldots+x_{q-1}+1)^2$,
 $\tilde x_2 =(\theta x_2+x_1+x_3+\ldots x_{q-1}+1)^2$,
 $\tilde x_3 =(\theta x_3+x_1+x_2+\ldots x_{q-1}+1)^2$, \ldots,
 $\tilde x_{q-1}=(\theta x_{q-1}+x_1+x_2+\ldots x_{q-2}+1)^2$,
$\gamma=\left(x_1+x_2+x_3+\ldots+x_{q-1}+\theta\right)^2$.

С обеих частей каждых равенств из (6) вычтем 1:

$$\sqrt{x_1}-1=\frac{(\theta-1)(\tilde x_1-\gamma)}{\tilde x_1+\tilde x_2+\tilde x_3+\ldots+\tilde x_{q-1}+\theta\gamma}$$
$$\sqrt{x_2}-1=\frac{(\theta-1)(\tilde x_2-\gamma)}{\tilde x_1+\tilde x_2+\tilde x_3+\ldots+\tilde x_{q-1}+\theta\gamma}$$
$$\sqrt{x_3}-1=\frac{(\theta-1)(\tilde x_3-\gamma)}{\tilde x_1+\tilde x_2+\tilde x_3+\ldots+\tilde x_{q-1}+\theta\gamma}$$
$$\ldots \ldots \ldots \ldots \ldots \ldots \ldots $$
$$\sqrt{x_{q-1}}-1=\frac{(\theta-1)(\tilde x_{q-1}-\gamma)}{\tilde x_1+\tilde x_2+\tilde x_3+\ldots+\tilde x_{q-1}+\theta\gamma}.$$
Введем обозначение
$$L=\frac{\theta-1}{\tilde x_1+\tilde x_2+\tilde x_3+\ldots+\tilde x_{q-1}+\theta\gamma}$$
и перепишем последнюю систему уравнений
$$\sqrt{x_i}-1=L(\tilde x_i-\gamma), \ i=1,2,\ldots q-1.\eqno(7)$$
Вычислим разности $\tilde x_i-\gamma, \ i=1,2,\ldots ,q-1$
$$\tilde x_i-\gamma=(\theta-1)\cdot(x_i-1)\cdot\left((\theta+1)(x_i+1)+2\left(\sum_{i=1}^{q-1}x_i-x_i\right)\right)$$
и подставим в (7). Тогда после некоторых преобразований получим
$$(\sqrt{x_i}-1)\cdot\left[1-L(\theta-1)\cdot(\sqrt{x_i}+1)\cdot\left((\theta+1)(x_i+1)+
2\left(\sum_{i=1}^{q-1}x_i-x_i\right)\right)\right]=0.$$
Отсюда при $i=1,2,\ldots q-1$ имеем $x_i=1$ или
$$1-L(\theta-1)\cdot(\sqrt{x_i}+1)\cdot\left((\theta+1)(x_i+1)+
2\left(\sum_{i=1}^{q-1}x_i-x_i\right)\right)=0.\eqno(8)$$
Заметим, что решение $x_i=1, \ i=1,2,\ldots q-1$ соответствует трансляционно-инвариантной мере Гиббса. Поэтому рассмотрим случай, когда $x_i\neq1$. Перепишем систему уравнений (8)
$$\frac{1}{L}=(\theta-1)\cdot(\sqrt{x_i}+1)\cdot\left((\theta+1)(x_i+1)+
2\left(\sum_{i=1}^{q-1}x_i-x_i\right)\right), \ i=1,2,\ldots q-1.$$
Тогда при $i\neq j, \ i,j=1,2,\ldots q-1$ будем иметь
$$(\theta+1)\cdot(\sqrt{x_i}+1)\cdot(x_i+1)+2\cdot(\sqrt{x_i}+1)\cdot\left(\sum_{i=1}^{q-1}x_i-x_i\right)=$$
$$=(\theta+1)\cdot(\sqrt{x_j}+1)\cdot(x_j+1)+2\cdot(\sqrt{x_j}+1)\cdot\left(\sum_{i=1}^{q-1}x_i-x_j\right).$$
После некоторых преобразований получим
$$(\sqrt{x_i}-\sqrt{x_j})\left[(\theta+1)(x_i+x_j+1)+(\theta-1)(\sqrt{x_i}+\sqrt{x_j}+\sqrt{x_ix_j})+
2\left(\sum_{i=1}^{q-1}x_i-x_i-x_j\right)\right]=0.$$
Отсюда $x_i=x_j$ или
$$(\theta+1)(x_i+x_j+1)+(\theta-1)(\sqrt{x_i}+\sqrt{x_j}+\sqrt{x_ix_j})+
2\left(\sum_{i=1}^{q-1}x_i-x_i-x_j\right)=0.\eqno(9)$$
В случае $x_i=x_j$ мы имеем решение $x_i=x_j=1$, которое соответствует трансляционно-инвариантной мере Гиббса. Пусть $x_i\neq x_j$. Тогда из (9) получим
$$\theta(x_i+x_j+1+\sqrt{x_i}+\sqrt{x_j}+\sqrt{x_ix_j})+2(x_1+x_2+x_3+x_4+...+x_{q-1}-x_i-x_j)+$$
$$+x_i+x_j+1-(\sqrt{x_i}+\sqrt{x_j}+\sqrt{x_ix_j})=0.\eqno(10)$$
Докажем, что уравнение (10) не имеет решений. Для этого достаточно показать справедливость неравенства
$$x_i+x_j+1>\sqrt{x_i}+\sqrt{x_j}+\sqrt{x_ix_j}.$$
Введя обозначения $\sqrt{x_i}=s, \ \sqrt{x_j}=t$, получим следующее квадратное неравенство относительно $s$
$$s^2-(t+1)s+t^2-t+1>0,$$
дискриминант которого $D=-3(t-1)^2<0$ отрицательный при $t\neq 1$. Следовательно, система уравнений (4) имеет решения только вида $x_i=x_j$, т.е. $z=(x_1,\ldots,x_{q-1},y_1,\ldots,y_{q-1})\in I_1$. Значит, все $G_k^{(2)}$-периодические меры Гиббса являются трансляционно-инвариантными, а единственность $G_k^{(2)}$-периодической меры Гиббса следует из единственности трансляционно-инвариантной меры Гиббса для антиферромагнитной модели Поттса (см. \cite{R}). Теорема доказана.\

\textbf{Замечание 3.} Для антиферромагнитной модели Поттса единственной трансляционно-инвариантной мере Гиббса соответствует решение системы уравнений (4) вида $x_i=y_i=1$, $i=1,2,\ldots,q-1$.\

\section{Ферромагнитный случай.}\

В работе \cite{RK} доказана следующая теорема.

\textbf{Теорема 4.}\cite{RK} Для модели Поттса с нулевым внешним полем
при $k\geq 1, \ q=3, \ J>0$ все $G_k^{(2)}$-периодические меры Гиббса являются
трансляционно-инвариантными.

Следующая теорема обобщает утверждение Теоремы 4.

\textbf{Теорема 5.} Пусть $k\geq2, \ q\geq 3, \ J>0, \ \alpha=0$. Тогда
для модели Поттса все $G_k^{(2)}$-периодические меры Гиббса являются
трансляционно-инвариантными.

\textbf{Доказательство.} Рассмотрим разности $x_i-y_i, \ i=1,2,\ldots,q-1$ в системе уравнений (4):
$$x_i-y_i=\left({(\theta-1)y_i + \sum_{j=1}^{q-1}y_j+1\over \sum_{j=1}^{q-1}y_j+\theta}-{(\theta-1)x_i + \sum_{j=1}^{q-1}x_j+1\over \sum_{j=1}^{q-1}x_j+\theta}\right)\cdot A_i=$$
$$={A_i\over XY} \cdot \left((\theta^2-\theta)(y_i-x_i)+(\theta-1) \left(y_i\sum_{j=1}^{q-1}x_j-x_i\sum_{j=1}^{q-1}y_j\right)+(\theta-1)\left(\sum_{j=1}^{q-1}y_j-
\sum_{j=1}^{q-1}x_j\right)\right),$$
где
$$A_i=\left({(\theta-1)y_i + \sum_{j=1}^{q-1}y_j+1\over \sum_{j=1}^{q-1}y_j+\theta}\right)^{k-1}+\ldots+\left({(\theta-1)x_i + \sum_{j=1}^{q-1}x_j+1\over \sum_{j=1}^{q-1}x_j+\theta}\right)^{k-1},$$
$$X=\sum_{j=1}^{q-1}x_j+\theta, \ Y=\sum_{j=1}^{q-1}y_j+\theta, \ i=\overline{1,q-1}.$$
После некоторых преобразований получим
$$x_i-y_i={A_i(\theta-1)\over XY} \cdot \left(\theta(y_i-x_i)+ y_i\sum_{j=1}^{q-1}(x_j-y_j)+(y_i-x_i)\sum_{j=1}^{q-1}y_j+\sum_{j=1}^{q-1}y_j-
\sum_{j=1}^{q-1}x_j\right).$$
В результате получим систему уравнений вида
$$\left\{%
\begin{array}{ll}
   a_{11}(x_1-y_1)+a_{12}(x_2-y_2)+...+a_{1q-1}(x_{q-1}-y_{q-1})=0\\[3 mm]
   a_{21}(x_1-y_1)+a_{22}(x_2-y_2)+...+a_{2q-1}(x_{q-1}-y_{q-1})=0\\[3 mm]
   . \ . \ . \ . \ . \ . \ . \ . \ . \ . \ . \ . \ . \ . \ . \ . \ . \ . \ . \ . \ . \ . \ . \ . \ . \ . \ . \ . \ . \ . \ . \ . \ . \ . \ . \ . \ . \ . \ . \ . \ . \ . \\[3 mm]
   a_{q-11}(x_1-y_1)+a_{q-12}(x_2-y_2)+...+a_{q-1q-1}(x_{q-1}-y_{q-1})=0,\\[3 mm]
\end{array}%
\right.\eqno(11)$$\
где
$$a_{ii}=1+{A_i(\theta-1)(\theta+1+\sum_{j=1}^{q-1}y_j-y_i)\over XY}, \ a_{il}={(\theta-1)(1-y_i)A_i\over XY}, \ i\neq l, \ i,l=\overline{1,q-1}.$$

Как известно, эта система имеет нуловое решение, если
определитель
$$\mathbf{\det A}=
\left|%
\begin{array}{cccccc}
1+{A_1(\theta-1)(1+Y-y_1)\over XY}&{(\theta-1)(1-y_1)A_1\over XY} &....&  {(\theta-1)(1-y_1)A_1\over XY} \\
{(\theta-1)(1-y_2)A_2\over XY} & 1+{A_2(\theta-1)(1+Y-y_2)\over XY}  & ....  & {(\theta-1)(1-y_2)A_2\over XY} \\
... & ...     &  ...     & ... \\
... & ...     & ...      & ... \\
... & ...     & ...      & ... \\
{(\theta-1)(1-y_{q-1})A_{q-1}\over XY} & {(\theta-1)(1-y_{q-1})A_{q-1}\over XY}         & ...        & 1+{A_{q-1}(\theta-1)(1+Y-y_{q-1})\over XY} \\
\end{array}%
\right|
$$
отличен от нуля, где $\mathbf{A}-$ матрица данной системы.
Перепишем определитель матрицы $\mathbf{A}$:
$$\mathbf{\det A}=C
\left|%
\begin{array}{cccccc}
1+B_1&1 &....&  1 \\
1 & 1+B_2  & ....  & 1 \\
... & ...     &  ...     & ... \\
... & ...     & ...      & ... \\
... & ...     & ...      & ... \\
1 & 1         & ...        & 1+B_{q-1} \\
\end{array}%
\right|,
$$
где
$$C=\frac{A_1(\theta-1)(1-y_1)}{XY}\cdot\frac{A_2(\theta-1)(1-y_2)}{XY}\cdot\ldots \cdot
\frac{A_{q-1}(\theta-1)(1-y_{q-1})}{XY},$$
$$B_i=\frac{XY+A_i(\theta-1)Y}{A_i(\theta-1)(1-y_i)}.$$\
Покажем, что $\mathbf{\det A}\neq0$. Для этого докажем следующую лемму.

\textbf{Лемма 3.} (\cite{FS}, задача 323).
$$\begin{vmatrix}
$$ 1+a_1 & 1&1&...&1&1 \\ 1 & 1+a_2&1&...&1&1
\\1&1&1+a_3&...&1&1\\...&...&...&...&...&...
\\...&...&...&...&...&...\\...&...&...&...&...&...\\
1&1&1&...&1&1+a_n\\ $$\end{vmatrix}
=a_1a_2a_3...a_n\cdot\left(1+\frac{1}{a_1}+\frac{1}{a_2}+...+\frac{1}{a_n}\right).$$

\textbf{Доказательство.} Разложим определитель по первой строке
$$\begin{vmatrix}$$ 1+a_1 & 1&1&...&1&1 \\ 1 &
1+a_2&1&...&1&1
\\1&1&1+a_3&...&1&1\\...&...&...&...&...&...
\\...&...&...&...&...&...\\...&...&...&...&...&...\\
1&1&1&...&1&1+a_n\\ $$\end{vmatrix}=\begin{vmatrix}$$ 1&
1&1&...&1&1 \\ 1 & 1+a_2&1&...&1&1
\\1&1&1+a_3&...&1&1\\...&...&...&...&...&...
\\...&...&...&...&...&...\\...&...&...&...&...&...\\
1&1&1&...&1&1+a_n\\ $$\end{vmatrix}+$$
$$+\begin{vmatrix}$$ a_1 &
0&0&...&0&0 \\ 1 & 1+a_2&1&...&1&1
\\1&1&1+a_3&...&1&1\\...&...&...&...&...&...
\\...&...&...&...&...&...\\...&...&...&...&...&...\\
1&1 &1&...&1&1+a_n\\ $$\end{vmatrix}$$

Каждый элемент первой строки первого определителя в правой части этого равенства умножим на -1 и сложим с остальными строками:
$$\Delta_1=\begin{vmatrix}
$$ 1& 1&1&...&1&1 \\ 1 & 1+a_2&1&...&1&1
\\1&1&1+a_3&...&1&1\\...&...&...&...&...&...
\\...&...&...&...&...&...\\...&...&...&...&...&...\\
1&1&1&...&1&1+a_n\\ $$\end{vmatrix}=\begin{vmatrix}$$ 1&1
&1&...&1&1 \\ 0 & a_2&0&...&0&0
\\0&0&a_3&...&0&0\\...&...&...&...&...&...
\\...&...&...&...&...&...\\...&...&...&...&...&...\\
0&0&0&...&0&a_n\\
$$\end{vmatrix}.$$
Разложив полученный определитель по первому столбцу, получим $\Delta_1=a_2a_3a_4...a_n$.
А второй определитель разложим по первой строке:
$$\begin{vmatrix}$$ a_1 &
0&0&...&0&0 \\ 1 & 1+a_2&1&...&1&1
\\1&1&1+a_3&...&1&1\\...&...&...&...&...&...
\\...&...&...&...&...&...\\...&...&...&...&...&...\\
1&1 &1&...&1&1+a_n\\ $$\end{vmatrix}=a_1\cdot\begin{vmatrix}$$1+a_2
& 1&1&...&1&1 \\ 1 & 1+a_3&1&...&1&1
\\1&1&1+a_4&...&1&1\\...&...&...&...&...&...
\\...&...&...&...&...&...\\...&...&...&...&...&...\\
1&1 &1&...&1&1+a_n\\ $$\end{vmatrix}.$$
Для определителя правой части этого равенства повторим вышеуказанную операцию:
$$a_1\cdot\begin{vmatrix}$$1+a_2 & 1&1&...&1&1 \\ 1 &
1+a_3&1&...&1&1
\\1&1&1+a_4&...&1&1\\...&...&...&...&...&...
\\...&...&...&...&...&...\\...&...&...&...&...&...\\
1&1 &1&...&1&1+a_n\\
$$\end{vmatrix}=a_1\cdot\begin{vmatrix}$$ 1& 1&1&...&1&1
\\ 1 & 1+a_3&1&...&1&1
\\1&1&1+a_4&...&1&1\\...&...&...&...&...&...
\\...&...&...&...&...&...\\...&...&...&...&...&...\\
1&1&1&...&1&1+a_n\\ $$\end{vmatrix}+$$
$$+a_1\cdot\begin{vmatrix}$$ a_2 &
0&0&...&0&0 \\ 1 & 1+a_3&1&...&1&1
\\1&1&1+a_4&...&1&1\\...&...&...&...&...&...
\\...&...&...&...&...&...\\...&...&...&...&...&...\\
1&1 &1&...&1&1+a_n\\ $$\end{vmatrix}.$$
Подобно $\Delta_1$ получим, что первое слагаемое правой части равен
$a_1a_3a_4...a_n$. А для второго определителя правой части повторим выше указанный процесс и т.д.
В результате получим
$$\begin{vmatrix}
$$ 1+a_1 & 1&1&...&1&1 \\ 1 & 1+a_2&1&...&1&1
\\1&1&1+a_3&...&1&1\\...&...&...&...&...&...
\\...&...&...&...&...&...\\...&...&...&...&...&...\\
1&1&1&...&1&1+a_n\\ $$\end{vmatrix}
=$$
$$=a_2a_3a_4\ldots a_n+a_1a_3a_4\ldots a_n+\ldots +a_1a_2a_3\ldots a_{n-1}+a_1a_2a_3\ldots a_n=$$
$$=a_1a_2a_3\ldots a_n\cdot\left(1+\frac{1}{a_1}+\frac{1}{a_2}+\ldots +\frac{1}{a_n}\right).$$
Лемма доказана.

Используя Лемму 3, вычислим определитель матрицы $A$:
$$\mathbf{\det A}=\frac{1}{X^{q-1}Y^{q-1}}\cdot\left(XY+A_1(\theta-1)Y\right)\cdot\left(XY+A_2(\theta-1)
Y\right)\cdot \ldots \cdot\left(XY+A_{q-1}(\theta-1)
Y\right)\times$$
$$\times\left(1+\frac{A_1(\theta-1)(1-y_1)}{XY+A_1(\theta-1)Y}+
\frac{A_2(\theta-1)(1-y_2)}{XY+A_2(\theta-1)Y}+\ldots +
\frac{A_{q-1}(\theta-1)(1-y_{q-1})}{XY+A_{q-1}(\theta-1)Y}
\right).$$
Отсюда, раскрыв скобки, получим
$$\mathbf{\det A}=\frac{1}{X^{q-1}Y^{q-1}}\cdot P,$$
где
$$P=\left(XY+A_1(\theta-1)Y\right)\cdot\left(XY+A_2(\theta-1)
Y\right)\cdot \ldots \cdot\left(XY+A_{q-1}(\theta-1)
Y\right)+$$
$$+\left(XY+A_2(\theta-1)Y\right)\cdot\left(XY+A_3(\theta-1)
Y\right)\cdot \ldots \cdot\left(XY+A_{q-1}(\theta-1)
Y\right)\times$$
$$\times A_1(\theta-1)(1-y_1)+\left(XY+A_1(\theta-1)Y\right)\cdot\left(XY+A_3(\theta-1)
Y\right)\cdot \ldots\cdot$$
$$\times\left(XY+A_{q-1}(\theta-1)
Y\right)\cdot A_2(\theta-1)(1-y_2)+\ldots+\left(XY+A_1(\theta-1)Y\right)\times$$
$$\times\left(XY+A_2(\theta-1)Y\right)\cdot \ldots \cdot\left(XY+A_{q-2}(\theta-1)Y\right)A_{q-1}(\theta-1)(1-y_{q-1}).$$

Далее, сгруппируем выражение для $P$ по степеням $(XY)$:
$$P=(XY)^{q-1}+(XY)^{q-2}(\theta-1)\cdot
\left[Y\sum_{j=1}^{q-1}A_j+(1-y_1)A_1+(1-y_2)A_2+\ldots+(1-y_n)A_n\right]+$$
$$+(XY)^{q-3}(\theta-1)^2Y\cdot\left[Y(A_1A_2+\ldots+A_{q-2}A_{q-1})+\sum_{i=1}^{q-1}\left(\sum_{j=1}^{q-1}A_j-A_i\right)
A_i(1-y_i)\right]+$$
$$+\ldots+(XY)^{q-1-i}(\theta-1)^iY^{i-1}\cdot[Y(A_1A_2\ldots A_i+\ldots+A_{q-i}A_{q-i+1}\ldots A_{q-1})+$$
$$+(A_2A_3\ldots A_{i}+\ldots+ A_{q-i+1}\ldots A_{q-1})A_1(1-y_1)+(A_1A_3\ldots A_i+\ldots+A_{q-i+1}\ldots A_{q-1})A_2(1-y_2)+$$
$$+\ldots+(A_1A_2\ldots A_{i-1}+\ldots+A_{q-i}A_{q-i+1}\ldots A_{q-2})A_{q-1}(1-y_{q-1})]+\ldots+$$
$$+(\theta-1)^{q-1}Y^{q-2}\cdot\left[Y(A_1A_2\ldots A_{q-1})+A_1A_2\ldots A_{q-1}(1-y_1)+\ldots+A_1A_2\ldots A_{q-1}(1-y_{q-1})\right].$$
Покажем, что $P>0$. Учитывая $\theta>1, \ X>0, \ Y>0, \ A_i>0$, достаточно показать, что выражения в квадратных скобках положительные. Действительно, после некоторых преобразований можно получить, что выражение в первой квадратной скобке
$$Y\sum_{j=1}^{q-1}A_j+(1-y_1)A_1+(1-y_2)A_2+\ldots+(1-y_n)A_n=(\theta+1)\sum_{j=1}^{q-1}A_j+\sum_{j\neq k}y_jA_k>0,$$
выражение во второй квадратной скобке
$$Y(A_1A_2+\ldots+A_{q-2}A_{q-1})+\left(\sum_{j=1}^{q-1}A_j-A_1\right)A_1(1-y_1)+\ldots+
\left(\sum_{j=1}^{q-1}A_j-A_{q-1}\right)A_{q-1}(1-y_{q-1})=$$
$$=(\theta+1)(A_1A_2+\ldots+A_{q-2}A_{q-1})+y_1\sum_{j\neq
k \neq 1}A_jA_k+\ldots+y_{q-1}\sum_{j\neq k \neq {q-1}}A_jA_k>0,$$
выражение в $i$-ой квадратной скобке
$$Y(A_1A_2\cdot\ldots\cdot A_i+\ldots+A_{q-i}A_{q-i+1}\cdot\ldots\cdot A_{q-1})+(A_2A_3\cdot\ldots\cdot
A_{i}+\ldots+ A_{q-i+1}\ldots A_{q-1})A_1(1-y_1)+$$
$$+(A_1A_3\cdot\ldots\cdot A_i+\ldots+A_{q-i+1}\cdot...\cdot A_{q-1}) A_2(1-y_2)+\ldots+$$
$$+(A_1A_2\cdot\ldots\cdot A_{i-1}+\ldots+A_{q-i}A_{q-i+1}\cdot\ldots\cdot A_{q-2}) A_{q-1}(1-y_{q-1}))=$$
$$=(\theta+1)(A_1\cdot\ldots \cdot A_{i}+\ldots+A_{q-i}\cdot\ldots \cdot A_{q-1})+y_1(A_2\cdot\ldots \cdot A_{i+1}+\ldots+A_{q-i}\cdot\ldots \cdot A_{q-1})+$$
$$+\ldots+y_{q-1}(A_1\cdot\ldots\cdot A_{i}+\ldots+A_{q-i-1}\cdot\ldots \cdot A_{q-2})>0$$
и наконец, выражение в последней квадратной скобке
$$YA_1A_2\ldots A_{q-1}+A_1A_2\ldots A_{q-1}(1-y_1)+\ldots+A_1A_2\ldots A_{q-1}(1-y_{q-1})=
A_1A_2\ldots A_{q-1}(\theta+1)>0.$$
Значит, $P>0$. Следовательно, система уравнений (11) имеет решения только вида $x_i=y_i$, т.е. $z=(x_1,\ldots x_{q-1}, y_1,\ldots y_{q-1})\in I_3$. Тогда по Лемме 2 получим требуемое. Теорема доказана.

\textbf{Замечание 4.} Для ферромагнитной модели Поттса доказана, что существуют $2^{q}-1$ трансляционно-инвариантных мер Гиббса (см.\cite{KRK}).

\end{document}